\documentclass[a4paper,10pt]{article}
\usepackage{pdfpages}
\usepackage{amsmath,amssymb}
\usepackage{hyperref}
\usepackage{verbatim}
\newenvironment{Eqnarray}{\arraycolsep 0.14em\begin{eqnarray}}{\end{eqnarray}}
\def\beqa{\begin{Eqnarray}}
\def\eeqa{\end{Eqnarray}}
\def\beq{\begin{Eqnarray}}
\def\eeq{\end{Eqnarray}}
\newcommand{\no}{\nonumber}
\begin{document}
\begin{titlepage}

\vskip1.5cm
\begin{center}
{\Large \bf The flavor of a light charged Higgs} \\
\end{center}
\vskip0.2cm

\begin{center}
Nicolás Bernal$^1$, Marta Losada$^1$, Yosef Nir$^2$, Yogev Shpilman$^2$   
\end{center}
\vskip 8pt

\begin{center}
{ \it $^1$New York University Abu Dhabi,\\
PO Box 129188, Saadiyat Island, Abu Dhabi, United Arab Emirates\\
$^2$Department of Particle Physics and Astrophysics,\\
Weizmann Institute of Science, Rehovot 7610001, Israel} \vspace*{0.3cm}

\href{mailto:nicolas.bernal@nyu.edu}{\tt nicolas.bernal@nyu.edu}, \href{mailto:marta.losada@nyu.edu}{\tt marta.losada@nyu.edu},\\ \href{mailto:yosef.nir@weizmann.ac.il}{\tt yosef.nir@weizmann.ac.il}, \href{mailto:yogev.shpilman@weizmann.ac.il}{\tt yogev.shpilman@weizmann.ac.il}
\end{center}

\vglue 0.3truecm

\begin{abstract}
\noindent
The ATLAS Collaboration has recently reported a search for light-charged Higgs in $t\to H^+ b$ decay, with $H^+\to c\bar b$. An excess with a local significance of approximately $3\sigma$ is found at $m_{H^+}\approx130$ GeV, with a best-fit value of ${\rm BR}(t\to H^+b)\times {\rm BR}(H^+\to c\bar b)=(1.6\pm0.6)\times10^{-3}$. We study the implications of such a hypothetical signal in multi-Higgs doublet models. We take into account constraints from searches for other charged Higgs decays and from flavor-changing neutral current processes. Two Higgs doublet models with flavor structure dictated by natural flavor conservation (NFC), minimal flavor violation (MFV), or the Froggatt-Nielsen (FN) mechanism cannot account for such excess. A three-Higgs doublet model with NFC can account for the signal. The Yukawa couplings of the neutral pseudoscalar $A$ in the down sector, $\hat Y_A^D$, should be larger by a factor of $4-6$ compared to the corresponding Yukawa couplings of the Higgs $h$, $\hat Y_h^D$. We further present two minimal scenarios, one in which a single Yukawa coupling in the down sector, $(\hat Y_A^D)_{bb}$, gives the only significant contribution, and one in which two Yukawa couplings in the up sector, $(\hat Y_A^U)_{tt}$ and $(\hat Y_A^U)_{tc}$, give the only significant contributions, and we discuss possible tests of these scenarios.
\end{abstract}
\end{titlepage}

\section{Introduction}
The ATLAS Collaboration has recently reported a search for a light-charged Higgs boson in $t\to H^+ b$ decay, with $H^+\to c\bar b$ \cite{ATLAS:2023bzb}. The observed exclusion limits are consistently weaker than expected. The largest excess in data has a local significance of about $3\sigma$ for 
\beq\label{eq:atlasmh}
m_{H^\pm}=130\ {\rm GeV},
\eeq
with a global significance of about $2.5\sigma$. The corresponding best fit is measured to be
\beq\label{eq:atlasbr}
{\rm BR}(t\to H^+b)\times {\rm BR}(H^+\to c\bar b)=(1.6\pm0.6)\times10^{-3}.
\eeq
Assuming that indeed Eqs. (\ref{eq:atlasmh}) and (\ref{eq:atlasbr}) are realized in Nature, we study the implications within multi-Higgs doublet models (MHDMs). For a recent related study, see Ref. \cite{Akeroyd:2022ouy}.

In models that include several scalar doublets, the neutral scalar particles mediate flavor-changing neutral current (FCNC) processes at tree level. If these states, beyond the Higgs particle of the Standard Model (SM), are not very heavy,  and their couplings to the SM fermions are generic, {\it i.e.} of order one with no special flavor structure, then these contributions to FCNC processes violate the experimental bounds. (For a recent review, see Refs.~\cite{Branco:2011iw, Crivellin:2013wna}.) Thus, various mechanisms that do enforce special flavor structures have been suggested. The most commonly studied frameworks are the following:
\begin{itemize}
\item Natural flavor conservation (NFC) \cite{Glashow:1976nt} - each fermion sector couples to a single scalar doublet.
\item Minimal flavor violation (MFV) \cite{DAmbrosio:2002vsn} - the flavor $[U(3)]^5$ global symmetry of the kinetic terms is broken only by the three SM Yukawa matrices.
\item Froggatt-Nielsen mechanism (FN) \cite{Froggatt:1978nt} - an approximate $U(1)$ symmetry dictates the structure of the Yukawa matrices.
\item Alignment (A2HDM) \cite{Pich:2009sp} - each of the three Yukawa matrices is proportional to the corresponding mass matrix. (This is automatically the case in the three Higgs doublet model with NFC \cite{Grossman:1994jb}.) 
\end{itemize}

In this work, we explore the following questions:
\begin{itemize}
\item What will we learn about the Yukawa matrices of extra scalars if Eqs. (\ref{eq:atlasmh}) and (\ref{eq:atlasbr}) are indeed realized in Nature? In particular, what is the minimal set of Yukawa couplings that can account for these measurements?
\item Are there any other unavoidable experimental signals that follow from these Yukawa couplings and would test the possible interpretations of the data?
\item What are the implications for the various flavor models listed above?
\end{itemize}

The plan of this paper is as follows. In Section \ref{sec:notations}, we introduce our notation and formalism. In Sections~\ref{sec:thb} and \ref{sec:hcb}, we consider the constraints from, respectively, the decay $t\to H^+b$ and the decay $H^+\to c\bar b$. In Section \ref{sec:fcnc} we present the constraints from FCNC processes. Sections \ref{sec:mhdm} and \ref{sec:minimal} analyze the implications on MHDMs with, respectively, various flavor structures and minimal number of new Yukawa couplings. We summarize our results in Section~\ref{sec:con}. 
 
\section{Notations and Formalism}
\label{sec:notations}
We study models with two Higgs doublets, $\Phi_{1,2}(1,2)_{+1/2}$. The scalar potential is given by (we use the notation of Ref. \cite{Gunion:2002zf})
\beqa\label{eq:scapot}
V&=& m_{11}^2|\Phi_1|^2+m_{22}^2|\Phi_2|^2-[m_{12}^2\Phi_1^\dagger\Phi_2+{\rm h.c.}]\no\\
&+&\frac12\lambda_1|\Phi_1|^4+\frac12\lambda_2|\Phi_2|^4+\lambda_3|\Phi_1|^2|\Phi_2|^2
+\lambda_4(\Phi_1^\dagger\Phi_2)(\Phi_2^\dagger\Phi_1)\no\\
&+&\left[\frac12\lambda_5(\Phi_1^\dagger\Phi_2)^2
+\Phi_1^\dagger\Phi_2(\lambda_6|\Phi_1|^2+\lambda_7|\Phi_2|^2)+{\rm h.c.}\right].
\eeqa
For simplicity, we consider a CP conserving scalar potential, in which case the scalar spectrum consists of the light and heavy neutral CP-even states, $h$ and $H^0$, the neutral CP-odd state $A$, and the charged Higgs $H^\pm$.

The Yukawa interactions are given by (we use the notation of Ref. \cite{Dery:2013aba})
\beq\label{eq:yukawa3}
{\cal L}_Y=\sum_{i=1,2}\left(\overline{Q}\widetilde\Phi_i Y^U_i {U}
+\overline{Q}\Phi_i Y^D_i {D}+\overline{L}\Phi_i Y^E_i {E}\right).
\eeq
We express the fermion mass matrices by dimensionless matrices $Y_M^F$:
\beq
Y_M^F=\frac{\sqrt2 M^F}{v},\ \ \ (F=U,D,E).
\eeq
We denote by $Y_S^F$, $S=h$, $H$, and $A$ the Yukawa matrices of these neutral physical scalars. We have:
\beqa
Y_M^F&=&+c_\beta Y_1^F+s_\beta Y_2^F,\no\\
Y_A^F&=&-s_\beta Y_1^F+c_\beta Y_2^F,\no\\
Y_h^F&=&-s_\alpha Y_1^F+c_\alpha Y_2^F,\no\\
Y_H^F&=&+c_\alpha Y_1^F+s_\alpha Y_2^F,
\eeqa
where
\beq
\tan\beta\equiv v_2/v_1,
\eeq
$c_\beta\equiv\cos\beta$, $s_\beta\equiv\sin\beta$, $c_\alpha\equiv\cos\alpha$, and $s_\alpha\equiv\sin\alpha$.
For our purposes, it will be useful to have the following relations:
\beqa\label{eq:yayh}
c_{\alpha-\beta}Y_A^F&=&s_{\alpha-\beta}Y_M^F+Y_h^F,\no\\
c_{\alpha-\beta}Y_H^F&=&Y_M^F+s_{\alpha-\beta}Y_h^F,\no\\
Y_H^F&=&s_{\alpha-\beta}Y_A^F+c_{\alpha-\beta}Y_M^F.
\eeqa

The charged Higgs mass eigenstate is given by
\beq
H^+=c_\beta h_2^+-s_\beta (h_1^-)^*.
\eeq
The charged Higgs couplings are given by the $Y_A^F$ couplings:
\beq\label{eq:yhplus}
{\cal L}_{H^\pm}=-\overline{D_L}H^- Y_A^U U_R
-\overline{U_L}H^+ Y_A^D D_R-\overline{\nu_L}H^+ Y_A^E E_R+{\rm h.c.}.
\eeq
The fermion mass matrices are given by
\beq\label{eq:ymass}
{\cal L}_{\rm mass}=-\overline{U_L}(v/\sqrt2) Y_M^U U_R
-\overline{D_L}(v/\sqrt2) Y_M^D D_R-\overline{E_L}(v/\sqrt2) Y_M^E E_R+{\rm h.c.}.
\eeq
We now transform to the fermion mass basis. The diagonalizing matrices $V_{fM}$, obtained via
\beqa
V_{uL}Y_M^U V_{uR}^\dagger&=&{\rm diag}(y_u,y_c,y_t)\equiv\hat Y^U_M,\no\\
V_{dL}Y_M^D V_{dR}^\dagger&=&{\rm diag}(y_d,y_s,y_b)\equiv\hat Y^D_M,\no\\
V_{eL}Y_M^E V_{eR}^\dagger&=&{\rm diag}(y_e,y_\mu,y_\tau)\equiv\hat Y^E_M,
\eeqa
define the transformation from the interaction to the mass basis:
\beqa
&&U^M=(u,c,t):\ \ \ U_L^M=V_{uL}U_L,\ \ \ U_R^M=V_{uR}U_R,\no\\
&&D^M=(d,s,b):\ \ \ D_L^M=V_{dL}D_L,\ \ \ D_R^M=V_{dR}D_R,\no\\
&&E^M=(e,\mu,\tau):\ \ \ E_L^M=V_{eL}E_L,\ \ \ E_R^M=V_{eR}E_R,
\eeqa
and the CKM matrix,
\beq
V=V_{uL}V_{dL}^\dagger.
\eeq
We can now write the Yukawa couplings of the charged Higgs (\ref{eq:yhplus}) in the fermion mass basis:
\beq\label{eq:yhplusm}
{\cal L}_{H^\pm}=-\overline{D_L^M}H^- V^\dagger \hat Y_A^U U_R^M
-\overline{U_L^M}H^+ V \hat Y_A^D D_R^M-\overline{\nu_L}H^+ \hat Y_A^E E_R^M+{\rm h.c.},
\eeq
where
\beq
\hat Y_A^U=V_{uL}Y_A^U V_{uR}^\dagger,\ \ \
\hat Y_A^D=V_{dL}Y_A^D V_{dR}^\dagger,\ \ \
\hat Y_A^E=V_{eL}Y_A^E V_{eR}^\dagger.
\eeq
We finally define the ratio of a diagonal entry in $\hat Y^F_A$ to the corresponding SM Yukawa coupling:
\beq\label{eq:defxif}
\xi_f\equiv(\hat Y^F_A)_{ff}/y_f,
\eeq
where $y_f=\sqrt{2}m_f/v$. For the fermion masses at the scale of $m_{H^+} = 130$ GeV, we use \cite{Huang:2020hdv}
%
\beqa
(m_u,m_c,m_t)&=&(1.2\times10^{-3},0.6,165)\ {\rm GeV},\no\\
(m_d,m_s,m_b)&=&(2.6\times10^{-3},0.052,2.77)\ {\rm GeV},\no\\
(m_e,m_\mu,m_\tau)&=&(5\times10^{-4},0.1,1.73)\ {\rm GeV},
\eeqa
so, in particular, $y_b=0.016$ and $y_t=0.95$. These values are used in section \ref{sec:fcnc} for a 1-loop contribution of the charged Higgs to FCNC processes. For the tree-level branching ratios, the SM pole masses are used. 

\section{The $\boldsymbol{t\to H^+b}$ decay}
\label{sec:thb}
We define $x_{it}\equiv m_i^2/m_t^2$, and specifically $x_{Ht}\equiv m_{H^+}^2/m_t^2$. In the following, we approximate $x_{bt}\approx0$.
The decay rate of $t\to H^+ b$ in the most general 2HDM is given by
\beq
 \Gamma(t\to H^+b)=\frac{m_t(1-x_{Ht})^2}{32\pi}\left[|(\hat Y_A^{U\dagger}V)_{tb}|^2+|(V\hat Y_A^{D})_{tb}|^2\right].
\eeqa
The rate of the dominant top decay mode, $t\to W^+ b$, is given by
\beqa
 \Gamma(t\to W^+b)&=&
\frac{g^2|V_{tb}|^2m_t(1-x_{Wt})}{64\pi}\left[1+1/x_{Wt}-2x_{Wt}\right].
\eeqa
Neglecting the $t\to H^+b$ contribution to the total decay rate, we obtain:
\beqa\label{eq:brttothb}
{\rm BR}(t\to H^+b)&=&
\frac{2(1-x_{Ht})^2\left[|(\hat Y_A^{U\dagger}V)_{tb}|^2+|(V\hat Y_A^{D})_{tb}|^2\right]}
{g^2|V_{tb}|^2(1-x_{Wt})\left[1+1/x_{Wt}-2x_{Wt}\right]}.
\eeqa
In order to satisfy the experimental result, Eq.~(\ref{eq:atlasbr}), we obviously need
\beq\label{eq:thblower}
{\rm BR}(t\to H^+b)\geq1\times10^{-3},
\eeq
where we took the $1\sigma$ lower bound. Putting in Eq. (\ref{eq:brttothb}) the following values of parameters
\beq
x_{Ht}=0.56,\ \ x_{Wt}=0.21,\ \ g=0.64,\ \ |V_{tb}|=1,
\eeq
we obtain
\beq\label{eq:brttohbnum}
{\rm BR}(t\to H^+b)=0.22\times\left[|(\hat Y_A^{U\dagger}V)_{tb}|^2+|(V\hat Y_A^{D})_{tb}|^2\right],
\eeq
which translates into
\beq
|(\hat Y_A^{U\dagger}V)_{tb}|\ {\rm and/or}\ |(V\hat Y_A^{D})_{tb}|\geq0.067.
\eeq

\begin{figure}[!t]
    \def\sepf{0.51}
    \centering
    \includegraphics[scale=\sepf]{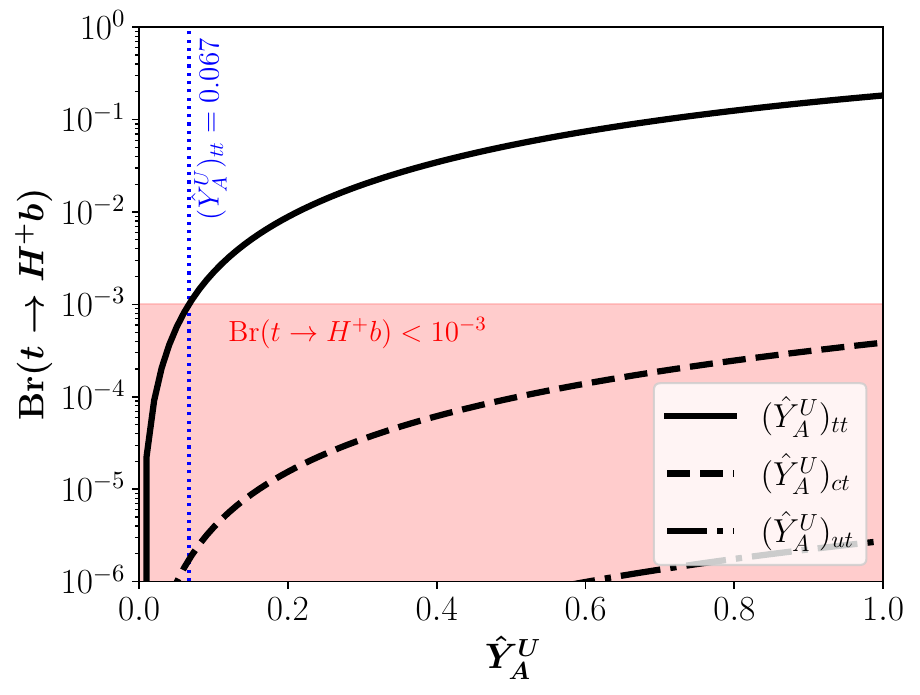}
    \includegraphics[scale=\sepf]{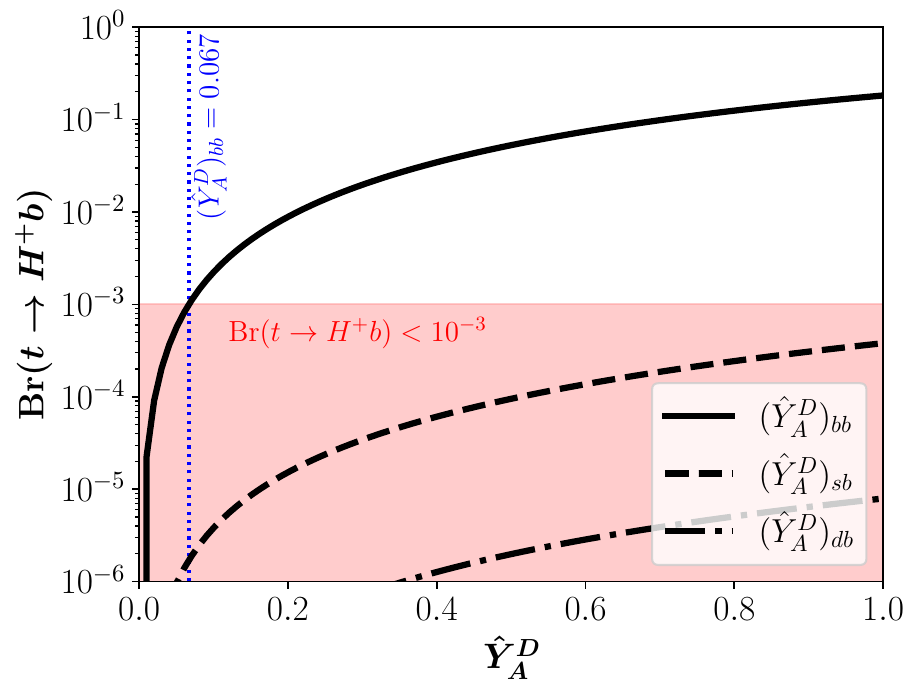}
    \caption{BR$(t\to H^+ b)$ for a single non-vanishing entry in $\hat Y^Q_A$: (left) $(\hat Y^U_A)_{it}$, and (right)  $(\hat Y^D_A)_{ib}$.}
    \label{fig:Brt2Hb}
\end{figure}

Taking into account that
\beq
|V_{tb}|\simeq1,\ \ |V_{cb}|\approx0.04,\ \ |V_{ub}|\approx0.004,
\eeq
and assuming that the entries of $\hat Y_A^U$ are $\lesssim 1$, we learn that $(\hat Y_A^U)_{ct}$ and $(\hat Y_A^U)_{ut}$ cannot play a dominant role in accounting for Eq.~(\ref{eq:thblower}).
In addition, noting that
\beq
|V_{ts}|\approx0.05,\ \ |V_{td}|\approx0.01,
\eeq
and assuming that the entries of $\hat Y_A^D$ are $\lesssim 1$, we learn that also $(\hat Y_A^D)_{sb}$ and $(\hat Y_A^D)_{db}$ cannot play a dominant role in accounting for Eq.~(\ref{eq:thblower}). Thus, we are led to the Yukawa alignment limit of 2HDM \cite{Pich:2009sp}, in which the off-diagonal elements in $\hat Y^A_F$ vanish, and we have
\beqa\label{eq:brttohbalign}
{\rm BR}(t\to H^+b)&=&
\frac{2(1-x_{Ht})^2\left[|(\hat Y_A^{U})_{tt}|^2+|(\hat Y_A^{D})_{bb}|^2\right]}
{g^2(1-x_{Wt})\left[1+1/x_{Wt}-2x_{Wt}\right]},
\eeqa
where the CKM dependence cancels out since both the numerator and the denominator depend on the leading CKM element $|V_{tb}|^2$. We need
\beq \label{eq:YA or YD}
|(\hat Y_A^{U})_{tt}|\ {\rm and/or}\ |(\hat Y_A^{D})_{bb}|\geq0.067.
\eeq
For more details, see Fig.~\ref{fig:Brt2Hb}.
If ${\rm BR}(H^+\to c\bar b)=1$, there is also an upper bound:
\beq \label{eq:uperyttybb}
|(\hat Y_A^{U})_{tt}|\ {\rm and}\ |(\hat Y_A^{D})_{bb}|\leq0.10.
\eeq

\section{The $\boldsymbol{H^+\to c\bar b}$ decay}
\label{sec:hcb}
In the following, we neglect $m_q/m_{H^+}$ for all quarks lighter than $H^+$. The decay rate of $H^+\to u_i \bar d_j$ (with $u_i=u$, $c$ and $d_j=d$, $s$, $b$) in the most general 2HDM is given by
\beq
 \Gamma(H^+\to u_i\bar d_j) = \frac{3m_{H^+}}{16\pi}\left(1+\frac{17}{3}\frac{\alpha_s(m_{H^+})}{\pi}\right)
\left[\left|(\hat Y_A^{U\dagger}V)_{ij}\right|^2+\left|(V\hat Y_A^{D})_{ij}\right|^2\right].
\eeq
Specifically, for $H^+\to c\bar b$, we have
\beq\label{eq:htocb}
\Gamma(H^+\to c\bar b) = \frac{3m_{H^+}}{16\pi}\left(1+\frac{17}{3}\frac{\alpha_s(m_{H^+})}{\pi}\right)
\left[\left|(\hat Y_{A}^{U})_{ic}V_{ib}\right|^2+\left|V_{ci}(\hat Y_{A}^{D})_{ib}\right|^2\right].
\eeq
The decay rate of $H^+\to \ell_i^+\nu_j$ in the most general 2HDM is given by
\beqa
 \Gamma(H^+\to \ell^+_i\nu_j)&=& \frac{m_{H^+}}{16\pi}\left|(\hat Y_A^{E\dagger})_{ij}\right|^2.
\eeqa

Assuming that the entries of $\hat Y_A^{U,D}$ are $\lesssim 1$, we learn from Eq. (\ref{eq:brttohbnum}) that ${\rm BR}(t\to H^+b)\lesssim 0.2$. Then, to satisfy the experimental result, Eq.~(\ref{eq:atlasbr}), we need
\beq\label{eq:hcblower}
{\rm BR}(H^+\to c\bar b)\gtrsim5\times10^{-3}.
\eeq
Experiments have put upper bounds on relevant decay rates \cite{CMS:2020osd},
\beq\label{eq:atlasbrcq}
{\rm BR}(t\to H^+b)\times {\rm BR}(H^+\to c\bar b+c\bar s)\leq2.7\times10^{-3},
\eeq
and \cite{ATLAS:2018gfm,CMS:2019bfg}
\beq\label{eq:atlasbrln}
{\rm BR}(t\to H^+b)\times {\rm BR}(H^+\to \tau^+\nu_j)\leq1.5\times10^{-3}.
\eeq
Then, we must also demand
\beqa\label{eq:cbcs}
\frac{\Gamma(H^+\to c\bar b)}{\Gamma(H^+\to c\bar s)}&=&\frac{|(\hat Y_{A}^{U})_{ic}V_{ib}|^2+|V_{ci}(\hat Y_{A}^{D})_{ib}|^2}
{|(\hat Y_{A}^{U})_{ic}V_{is}|^2+|V_{ci}(\hat Y_{A}^{D})_{is}|^2}\gtrsim 0.6,\\
\frac{\Gamma(H^+\to c\bar b)}{\Gamma(H^+\to \tau^+\nu_j)}&\approx&\frac{3(|(\hat Y_{A}^{U})_{ic}V_{ib}|^2+|V_{ci}(\hat Y_{A}^{D})_{ib}|^2)}
{|(\hat Y_{A}^{E})_{j\tau}|^2}\gtrsim 0.7.\label{eq:cbtn}
\eeqa

\begin{figure}[!t]
    \def\sepf{0.51}
    \centering
    \includegraphics[scale=\sepf]{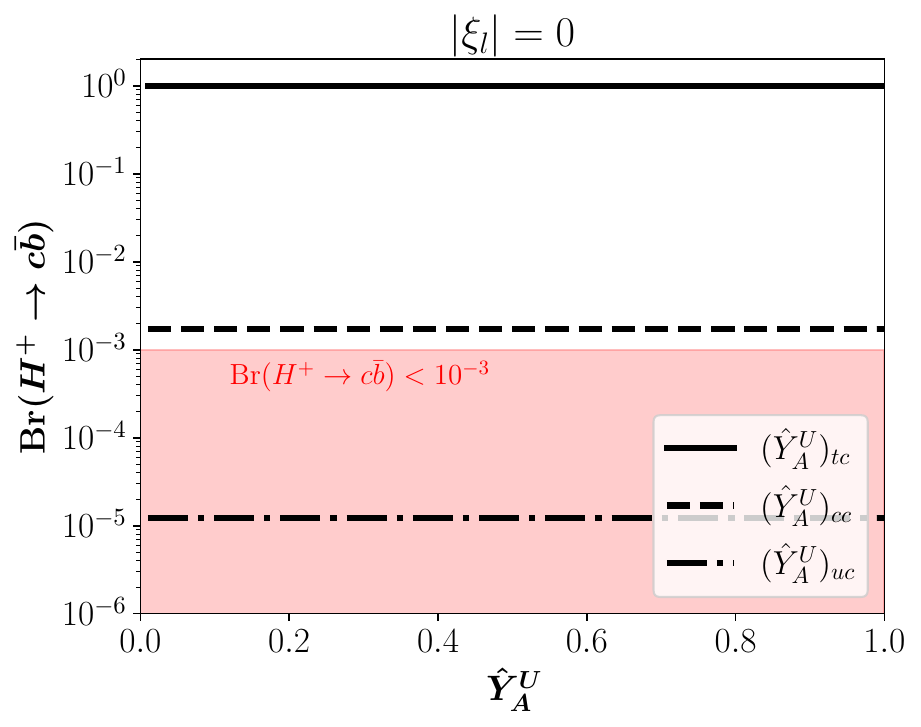}
    \includegraphics[scale=\sepf]{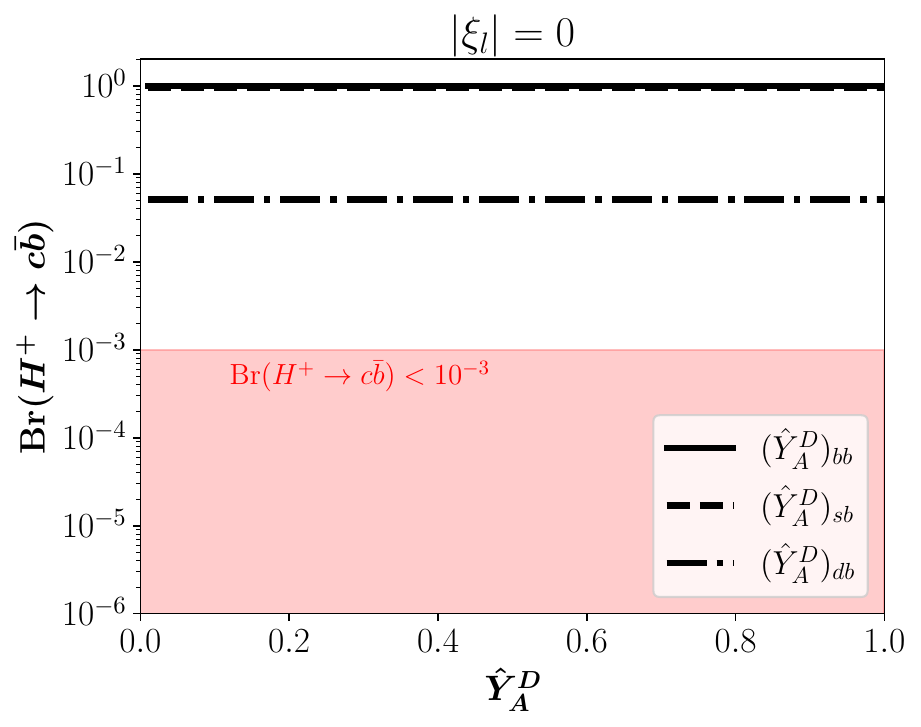}
    \caption{BR$(H^+ \to c \bar b)$ for a single non-vanishing entry in $\hat Y^Q_A$: (left) $(\hat Y^U_A)_{ic}$, and (right)  $(\hat Y^D_A)_{ib}$.}
    \label{fig:BrH2cb}
\end{figure}

Let us examine the various scenarios where $\Gamma(H^+\to c\bar b)$ is dominated by a single entry in $\hat Y_A^Q$:
\begin{itemize}
\item $(\hat Y^U_A)_{tc}$: we can have ${\rm BR}(H^+\to c\bar b)\simeq1$, provided that $|(\hat Y_{A}^{E})_{j\tau}|\lesssim 2|(\hat Y^U_A)_{tc}|$.
\item $(\hat Y^U_A)_{cc}$: we have $\Gamma(H^+\to c\bar b)/\Gamma(H^+\to c\bar s)\approx|V_{cb}/V_{cs}|^2\sim1.6\times10^{-3}$, which violates Eq.~(\ref{eq:cbcs}).
\item $(\hat Y^U_A)_{uc}$: ${\rm BR}(H^+\to c\bar b)\leq|V_{ub}/V_{ud}|^2\sim1.6\times10^{-5}$, which violates Eq.~(\ref{eq:hcblower}).
\item $(\hat Y^D_A)_{bb}$: we can have ${\rm BR}(H^+\to c\bar b)\simeq1$, provided that $|(\hat Y^D_A)_{bs}|\lesssim 1.3|(\hat Y^D_A)_{bb}|$ and $|(\hat Y_{A}^{E})_{j\tau}|\lesssim 2|(\hat Y^D_A)_{bb}V_{cb}|$. For $(\hat Y^D_A)_{bb}$ the only nonvanishing coupling, we have ${\rm BR}(H^+\to u\bar b)\simeq|V_{ub}/V_{cb}|^2\sim0.01$.
\item $(\hat Y^D_A)_{sb}$: we can have ${\rm BR}(H^+\to c\bar b)\simeq1$, provided that $|(\hat Y^D_A)_{ss}|\lesssim 1.3|(\hat Y^D_A)_{sb}|$ and $|(\hat Y_{A}^{E})_{j\tau}|\lesssim 2|(\hat Y^D_A)_{sb}V_{cs}|$. For $(\hat Y^D_A)_{sb}$ the only nonvanishing coupling, we have ${\rm BR}(H^+\to u\bar b)\simeq|V_{us}/V_{cs}|^2\sim0.05$.
\item $(\hat Y^D_A)_{db}$: we can have ${\rm BR}(H^+\to c\bar b)\simeq|V_{cd}/V_{ud}|^2\sim0.05$, provided that $|(\hat Y^D_A)_{ds}|\lesssim 1.3|(\hat Y^D_A)_{db}|$ and $|(\hat Y_{A}^{E})_{j\tau}|\lesssim 2|(\hat Y^D_A)_{db}V_{cd}|$. For $(\hat Y^D_A)_{db}$ the only nonvanishing coupling, we have ${\rm BR}(H^+\to u\bar b)\sim0.95$.
\end{itemize}
For more details, see Fig.~\ref{fig:BrH2cb} .

We learn the following lessons:
\begin{itemize}
\item $(\hat Y^U_A)_{cc}$ and $(\hat Y^U_A)_{uc}$ cannot play a dominant role in explaining the experimental result. 
\item A single entry,  $(\hat Y^D_A)_{bb}\sim0.07$, and all other entries in $\hat Y^U_A$ and $\hat Y^D_A$ negligibly small, can explain the experimental result.
\item The experimental result can be explained in the aligned limit, but requires $|(\hat Y^U_A)_{cc}/(\hat Y^D_A)_{bb}|\lesssim |V_{cb}/V_{cs}|\sim0.05$ and $|(\hat Y^E_A)_{\tau\tau}/(\hat Y^D_A)_{bb}|\lesssim 2|V_{cb}|\sim0.09$.
\end{itemize} 

\section{FCNC constraints}
\label{sec:fcnc}

The most relevant FCNC processes for our purposes are $b\to s\gamma$ decay, which constrains the product of a $\hat Y^U_A$ and a $\hat Y^D_A$ coupling, $D^0-\overline{D}{}^0$ mixing, where there is an unavoidable contribution in our minimal scenario with only $(\hat Y^D_A)_{bb}\neq0$, $t\to c\gamma$ decay, which also constrains the product of a $\hat Y^U_A$ and a $\hat Y^D_A$ coupling, and $B^0-\overline{B}{}^0$ mixing, where there is a contribution from $(\hat Y^U_A)_{tt}\neq0$. We discuss these four processes in the following.

Off-diagonal entries in $\hat Y^A_F$ contribute to FCNC mediated by the neutral scalars. Here, we do not consider these contributions and focus on the contributions of the charged Higgs to FCNC. There are several reasons why we can do so:
\begin{itemize}
\item In the NFC models, discussed in Section \ref{sec:mhdm}, the neutral scalars do not contribute to FCNC processes.
\item In the other flavor models discussed in Section \ref{sec:mhdm}, the flavor structure is close to NFC, and the off-diagonal couplings are highly suppressed.
\item Although the splitting between the mass of $H^\pm$ and those of the neutral scalars $A$ and $H^0$ is of order $v$, $A$ and $H^0$ can still be considerably heavier than $H^\pm$ (say, 500 GeV), providing mass-suppression to their contributions.
\item The contributions of the neutral scalars could be significant, even if they are a factor of a few heavier than the charged scalar, if they contributed at tree level to FCNC. This does not happen in radiative decays ($b\to s\gamma$ and $t\to c\gamma$), but could in principle happen for neutral-meson mixing. However, it happens in neither of our minimal (non-NFC) models, discussed in Section \ref{sec:minimal}.
\end{itemize}

\subsection{The $\boldsymbol{b\to s\gamma}$ decay}
The particle data group quotes the experimental value \cite{ParticleDataGroup:2022pth}
\beq
{\rm BR}(b\to s\gamma)^{\rm exp}_{E_\gamma>1.6\ {\rm GeV}}=(3.49\pm0.19)\times10^{-4}.
\eeq
The SM prediction is given by \cite{Misiak:2015xwa}
\beq
{\rm BR}(b\to s\gamma)^{\rm sm}_{E_\gamma>1.6\ {\rm GeV}}=(3.36\pm0.23)\times10^{-4}.
\eeq
This leaves the following window for new physics (and, in particular, charged Higgs) contribution:
\beq\label{eq:boundhbsg}
{\rm BR}(b\to s\gamma)^{H^+}_{E_\gamma>1.6\ {\rm GeV}}=(0.13\pm0.30)\times10^{-4}.
\eeq

The contribution of the charged Higgs-top loop is given by \cite{Trott:2010iz}
\beqa
{\rm BR}(b\to s\gamma)^{H^+t}_{E_\gamma>1.6\ {\rm GeV}}&=&-2.7\times10^{-4}\times[f_\gamma(x_{Ht}^{-1})/x_{Ht}]\times{\cal R}e(\xi_t\xi_b)\no\\
&=&-1.9\times10^{-4}\times{\cal R}e(\xi_t\xi_b),
\eeqa
where 
\beq
f_{\gamma}(z) =\frac{1}{4}\left(\frac{1+2z\ln z-z^{2}}{\left(1-z\right)^{3}}\right)-\left(\frac{1+\ln z-z}{\left(1-z\right)^{2}}\right),
\eeq
$\xi_f$ is defined in Eq.~(\ref{eq:defxif}), and the second equality corresponds to $m_{H^+}=130$~GeV. Thus, we obtain the $1\sigma$ range
\beq\label{eq:bsgtb}
-0.22\lesssim  {\cal R}e(\xi_t\xi_b)\lesssim +0.09.
\eeq
%

\subsection{The $\boldsymbol{t\to c\gamma}$ decay}
The ATLAS \cite{ATLAS:2022per} and CMS \cite{CMS:2023tir} collaborations have searched for the $t\to c\gamma$ decay, and quote upper bounds:
\beq
{\rm BR}(t\to c\gamma)<\left\{
\begin{matrix}4.2\times10^{-5} & {\rm ATLAS},\\
1.5\times10^{-5} & {\rm CMS}.\end{matrix}\right.
\eeq
The charged Higgs contribution to this decay is given by \cite{Desai:2022zig}
\beqa
{\rm BR}(t\to c\gamma)&=&\frac{m_t^5}{4\pi\Gamma_t}
\left(\frac{5e}{1152\pi^2m_{H^+}^2}\right)^2 (y_t y_b)^2|V_{tb}Y_{cb}^*|^2||\xi_t\xi_b|^2\no\\
&\approx&5.7\times10^{-7}(y_t y_b)^2|V_{tb}Y_{cb}^*|^2||\xi_t\xi_b|^2
\sim2.6\times10^{-13}|\xi_t\xi_b|^2.
\eeqa

We conclude that the charged Higgs contribution to the decay $t\to c\gamma$ is much below the experimental sensitivity.

\subsection{$\boldsymbol{D^0-\overline{D}{}^0}$ mixing}
The experimental value of the mass splitting between the neutral $D$-meson mass eigenstates is given by~\cite{HeavyFlavorAveragingGroup:2022wzx}
\beq
\Delta m_D/\Gamma_D=(4.1\pm0.5)\times10^{-3}.
\eeq
 The contribution of $(Y^D_A)_{bb}$ to $\Delta m_D$ can be extracted from the expressions in Ref. \cite{Barger:1989fj}:
\beqa
\Delta m_{D}^{H^+}&=&\frac{G_{F}^{2}M_{W}^{2}}{24\pi^{2}}m_{D}f_{D}^{2}B_{D}\left|V_{cb}^{*}V_{ub}\right|^{2}x_{bW}^2\\
&&\times\left\{\xi_b^{4} I_1\left(x_{bW},x_{HW}\right)+\xi_b^{2} \left[8I_3(x_{bW},x_{HW})-2I_4(x_{bW},x_{HW})\right]\right\},\no
\eeqa
where $\xi_b$ is defined in Eq.~(\ref{eq:defxif}), $x_{iW} =m_{i}^{2}/m_{W}^{2}$, we put $x_{cW}=0$, and
\begin{align}\label{eq:I134}
I_{1}\left(x,y\right) & =\frac{x+y}{\left(x-y\right)^{2}}-\frac{2xy}{\left(x-y\right)^{3}}\log\frac{x}{y}\\
I_{3}\left(x,y\right) & =\frac{1}{\left(x-y\right)\left(1-x\right)}+\frac{y\log y}{\left(x-y\right)^{2}\left(1-y\right)}+\frac{\left(x^{2}-y\right)\log x}{\left(x-y\right)^{2}\left(1-x\right)^{2}}\no\\
I_{4}\left(x,y\right) & =\frac{x}{\left(x-y\right)\left(1-x\right)}+\frac{y^{2}\log y}{\left(1-y\right)\left(x-y\right)^{2}}+\frac{x\left(x+xy-2y\right)\log y}{\left(x-y\right)^{2}\left(1-x\right)^{2}}.\no
\end{align}

We obtain (using $\tau_D=4.1\times10^{-13}$ s):
\begin{equation}
\Delta m_{D}^{H^+}/\Gamma_{D}\approx3\times10^{-10}\left(\xi_b\right)^{2}+7.6\times10^{-12}\left(\xi_b\right)^{4}.
\end{equation}
We conclude that $D^0-\overline{D}{}^0$ mixing requires, for $m_{H^+}=130$ GeV,
\beq
\xi_b\lesssim 90.
\eeq
%

\subsection{$\boldsymbol{B^0-\overline{B}{}^0}$ mixing}
The experimental value of the mass splitting between the neutral $B$-meson mass eigenstates is given by~\cite{HeavyFlavorAveragingGroup:2022wzx}
\beq
\Delta m_B/\Gamma_B=0.769\pm0.004.
\eeq
It has been shown that new physics is constrained to contribute less than about 20 percent to the mixing amplitude \cite{Charles:2015gya}.
 The contribution of $(Y^U_A)_{tt}$ to $\Delta m_B$ can be extracted from the expressions in Ref. \cite{Barger:1989fj}:
\beqa
\Delta m_{B}^{H^+}&=&\frac{G_{F}^{2}M_{W}^{2}}{24\pi^{2}}m_{B}f_{B}^{2}B_{B}\left|V_{tb}^{*}V_{td}\right|^{2}x_{tW}^2\\
&&\times\left\{\xi_t^{4} I_1\left(x_{tW},x_{HW}\right)+\xi_t^{2} \left[8I_3(x_{tW},x_{HW})-2I_4(x_{tW},x_{HW})\right]\right\},\no
\eeqa
where $\xi_t$ is defined in Eq.~(\ref{eq:defxif}), we put $x_{bW}=0$, and $I_{1,3,4}$ are given in Eq. (\ref{eq:I134}).

We obtain (using $\tau_B=1.5\times10^{-12}$ s):
\begin{equation}
\Delta m_{B}^{H^+}/\Gamma_{B}\approx3.0\left(\xi_t\right)^{2}+0.26\left(\xi_t\right)^{4}.
\end{equation}
We conclude that $B^0-\overline{B}{}^0$ mixing requires, for $m_{H^+}=130$ GeV,
\beq
\xi_t\lesssim 0.2.
\eeq

Importantly, the charged Higgs contribution to the $B^0-\overline{B}{}^0$ mixing amplitude carries the same phase as the SM contribution. In particular, this contribution depends on $|(Y^U_A)_{tt}|$ and is independent of ${\rm arg}[(Y^U_A)_{tt}]$. Thus, CP asymmetries in neutral $B$ decays, such as $S_{\psi K_S}$, are not affected.

\subsection{Summary}
\begin{figure}[!t]
    \def\sepf{0.50}
    \centering
   \includegraphics[scale=\sepf]{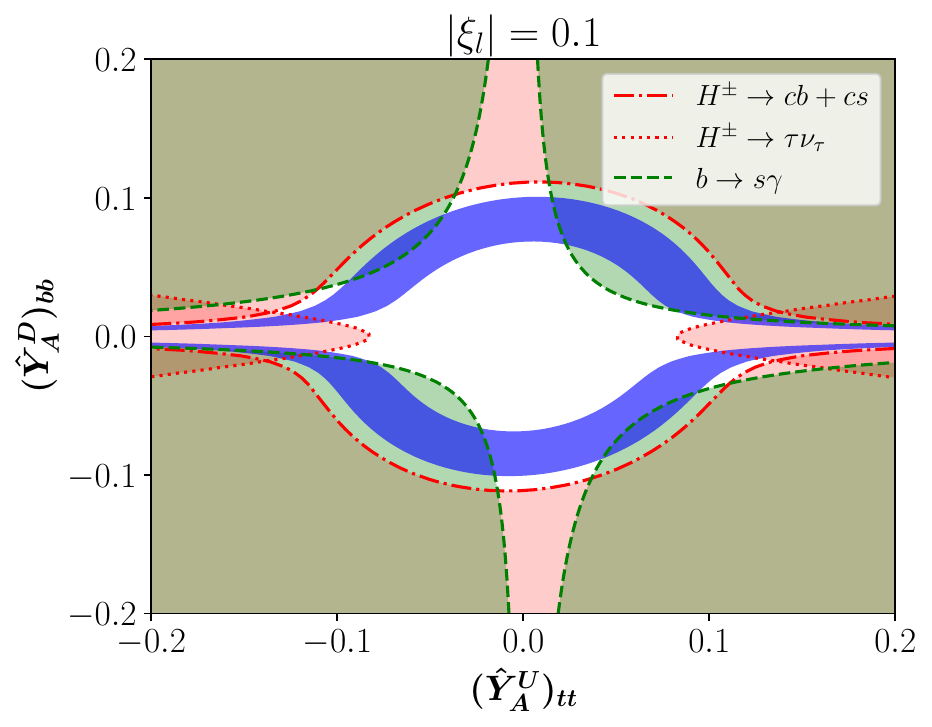}
   \includegraphics[scale=\sepf]{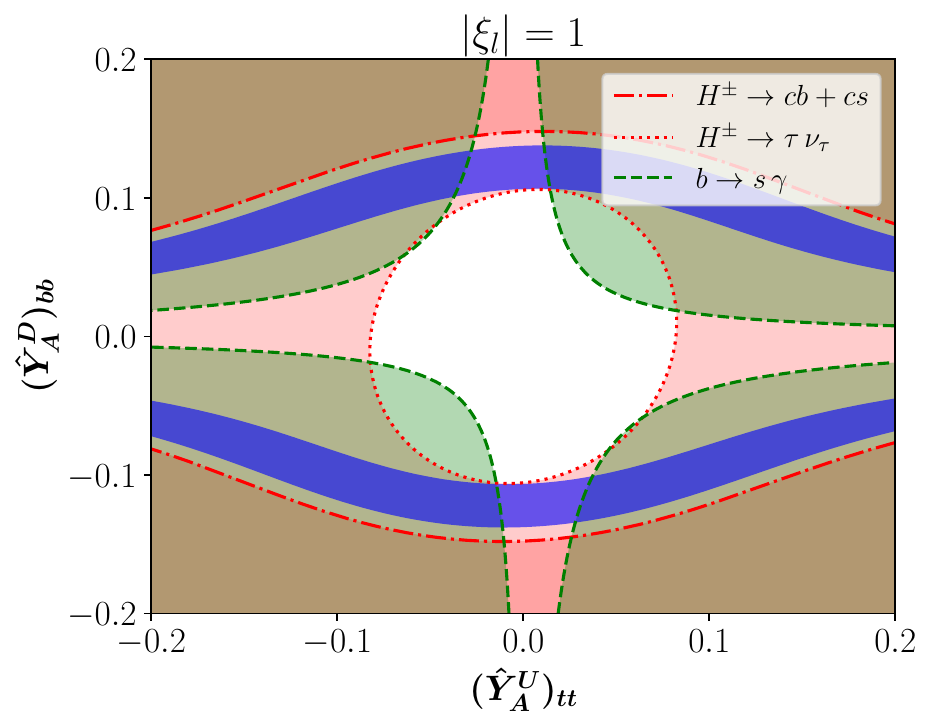}
    \caption{Constraints in the $(\hat Y^U_A)_{tt}-(\hat Y^D_A)_{bb}$ plane in the case that the only significant $(\hat Y^F_A)$ couplings are to the third generation fermions, for two values of $(\hat Y^E_A)_{\tau\tau}/y_\tau$: (left) 0.1, (right) 1.0. The blue region corresponds to ${\rm BR}(t\to H^+ b)\times{\rm BR}(H^+\to c\bar b)$ of Eq.~(\ref{eq:atlasbr}). The green (outside the dashed curves), red (outside the dot-dashed curves), and light red (outside the dotted curves) regions are excluded by the branching ratios of, correspondingly, $b\to s\gamma$ [Eq.~(\ref{eq:bsgtb})], $H^+\to c\bar b+c\bar s$  [Eq.~(\ref{eq:cbcs})] and $H^+\to\tau^+\nu$  [Eq.~(\ref{eq:cbtn})].}
     \label{fig:xitxib}
\end{figure}

The most significant exclusion limits on the $H^+$ couplings come from the upper bounds on ${\rm BR}(H^+\to c\bar b+c\bar s)$ and ${\rm BR}(H^+\to\tau\nu_j)$ [see Eq.~(\ref{eq:cbcs})], and from the allowed room for a contribution of new physics to ${\rm BR}(b\to s\gamma)$ [see Eq.~(\ref{eq:bsgtb})]. We present these bounds for the representative case where the only significant couplings of $H^+$ are to the third-generation fermions in Fig.~\ref{fig:xitxib}, consistent with the numerical analysis of previous sections.

Before closing this section, we note that in the case where $\xi_\ell \ne 0$, further constraints from the leptonic sector can arise. However, here they are not discussed, as $Y^E_A$ couplings are not required for the explanation of the excess.

\section{Flavor models}
\label{sec:mhdm}
\subsection{2HDM with NFC}
The basic principle of multi-Higgs doublet models (MHDM) with natural flavor conservation (NFC) is that each fermion sector ($U$, $D$, $E$) couples to only one Higgs doublet. The Yukawa Lagrangian is given by
\beq\label{eq:yukawanfc}
{\cal L}_Y=\overline{Q}\widetilde\Phi_u Y^U U+\overline{Q} \Phi_d Y^D D+\overline{L}\Phi_e Y^E E.
\eeq
In two Higgs doublet models (2HDM), by convention, $\Phi_u=\Phi_2$, while each of $\Phi_d$ and $\Phi_e$ can be either $\Phi_2$ or $\Phi_1$. In these models,
\beq\label{eq:yaymnfc}
\hat Y^F_A=\xi_F\hat Y^F_M, 
\eeq
yielding four different NFC models:
\begin{itemize}
\item Type I: $\xi_U=\xi_D=-\xi_E$.
\item Type II: $\xi_U=1/\xi_D=1/\xi_E$.
\item Type III: $\xi_U=-\xi_D=1/\xi_E$.
\item Type IV: $\xi_U=1/\xi_D=-\xi_E$.
\end{itemize}

The contributions of the charged Higgs to some of the observables that we considered become independent of $\xi_F$ in specific NFC types:
\begin{itemize}
\item $b\to s\gamma$, for $m_{H^+}=130$ GeV:  
\beq
{\rm BR}(b\to s\gamma)^{H^+t}_{E_\gamma>1.6\ {\rm GeV}}=-1.9\times10^{-4}\ \ ({\rm Type\ II\ and\ Type\ IV}),
\eeq
in contradiction to Eq.~(\ref{eq:boundhbsg}).
\item $H^+\to c\bar b+c\bar s$:
\beq
\frac{\Gamma(H^+\to c\bar b)}{\Gamma(H^+\to c\bar s)}\approx\left|\frac{V_{cb}}{V_{cs}}\right|^2\frac{m_b^2}{m_c^2}\sim0.09\ \ ({\rm Type\ I\ and\ Type\ III}),
\eeq
in contradiction to Eq.~(\ref{eq:cbcs}).
\item $H^+\to\tau^+\nu_\tau$:
\beq
\frac{\Gamma(H^+\to c\bar b)}{\Gamma(H^+\to \tau^+\nu_\tau)}&\approx&3|V_{cb}|^2\frac{m_b^2}{m_\tau^2}\sim0.013\ \ ({\rm Type\ I\ and\ Type\ II}),
\eeq
in contradiction to Eq.~(\ref{eq:cbtn}).
\end{itemize}
We conclude that, if Eqs.~(\ref{eq:atlasmh}) and (\ref{eq:atlasbr}) are realized in Nature, all four types of 2HDM-NFC models will be excluded.

\subsection{3HDM with NFC}
\label{sec:3hdmnfc}
With three or more Higgs doublets, the NFC Yukawa Lagrangian is still given by Eq.~(\ref{eq:yukawanfc}), resulting in the relations of Eq.~(\ref{eq:yaymnfc}). However, the possibility that $\Phi_u$, $\Phi_d$ and $\Phi_e$ are three different Higgs doublets, in which case $\xi_U$, $\xi_D$ and $\xi_E$ are three independent parameters, opens up. In this section, we discuss this specific NFC scenario. Note that in the 3HDM, there are two charged Higgs fields. We implicitly assume below that the charged Higgs relevant to Eqs. (\ref{eq:atlasmh}) and (\ref{eq:atlasbr}) is the lighter of the two.

\begin{figure}[!t]
    \def\sepf{0.52}
    \centering
    \includegraphics[scale=\sepf]{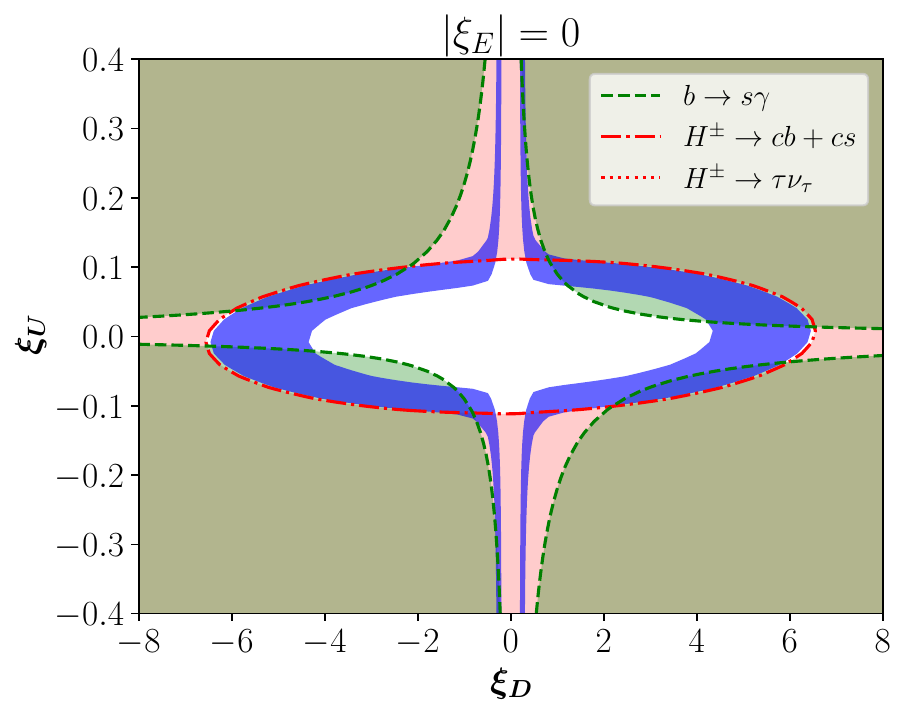}
    \includegraphics[scale=\sepf]{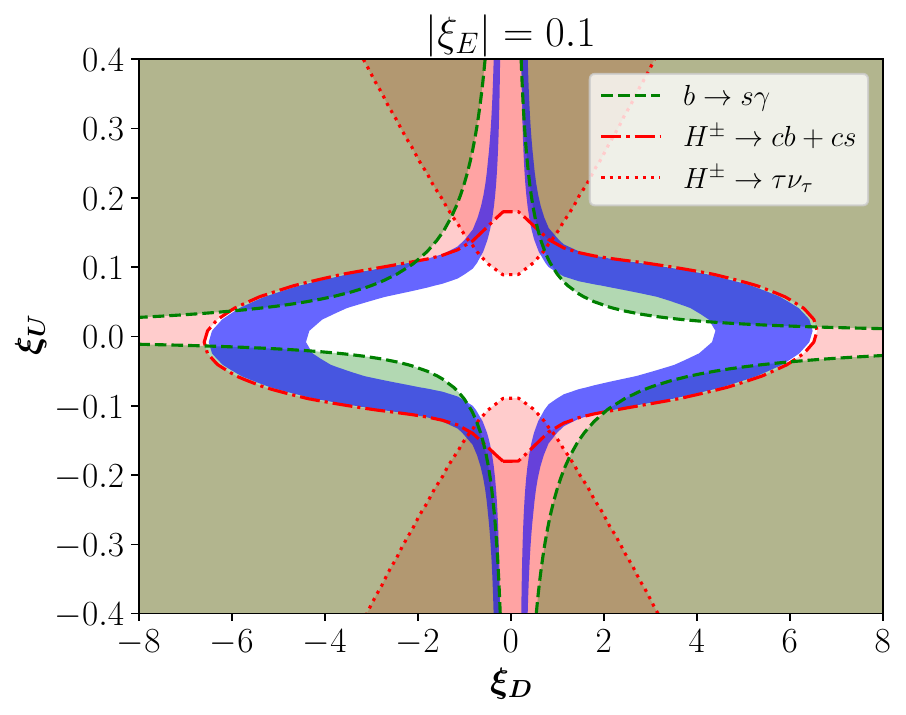}
    \includegraphics[scale=\sepf]{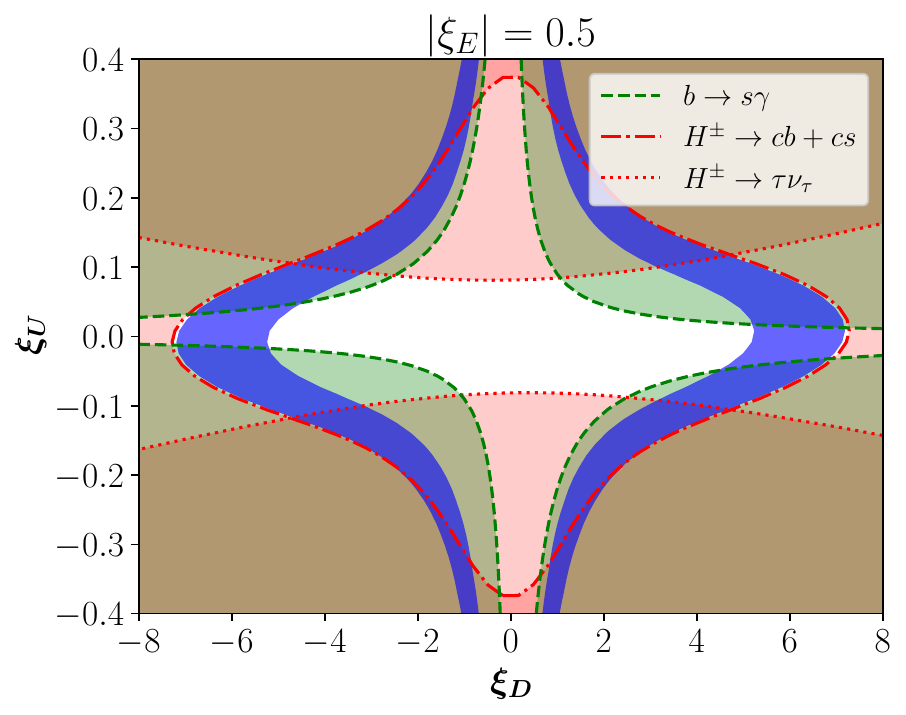}
    \includegraphics[scale=\sepf]{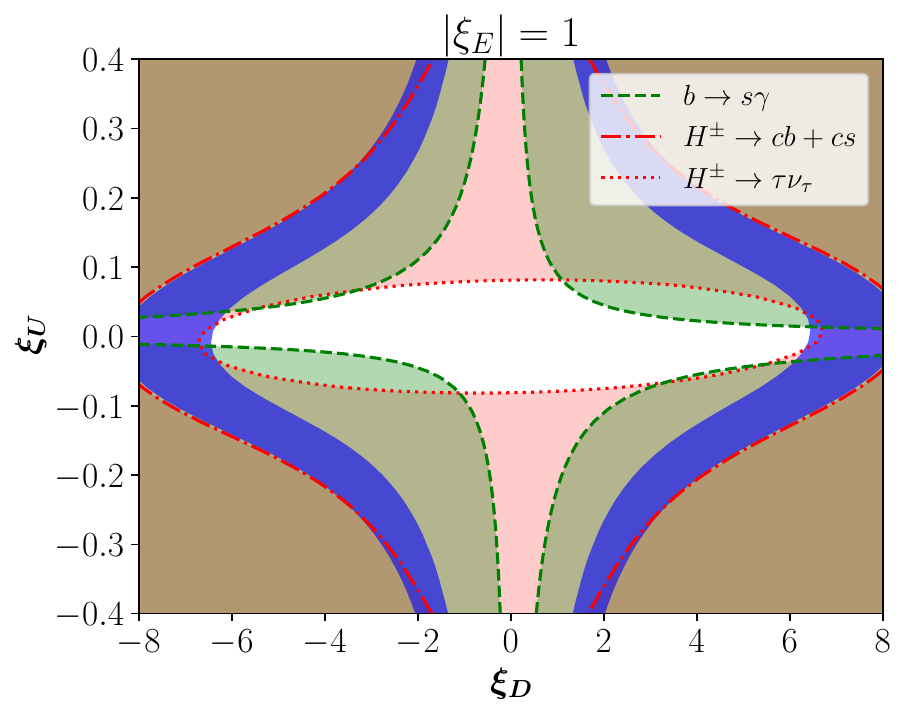}
    \caption{Constraints in the $\xi_U-\xi_D$ plane in the 3HDM-NFC model, where each fermion sector couples to a different Higgs doublet. The blue region corresponds to ${\rm BR}(t\to H^+ b)\times{\rm BR}(H^+\to c\bar b)$ of Eq.~(\ref{eq:atlasbr}). The green (outside the dashed curves), red (outside the dot-dashed curves), and light red (outside the dotted curves) regions are excluded by, correspondingly, BR$(b\to s\gamma)$ [Eq.~(\ref{eq:bsgtb})], BR$(H^+\to c\bar b+c\bar s)$  [Eq.~(\ref{eq:cbcs})] and BR$(H^+\to\tau^+\nu)$  [Eq.~(\ref{eq:cbtn})]. The four plots correspond to four different values of $\xi_E$: (upper left) $0$, (upper right) $0.1$, (lower left) $0.5$, and (lower right) $1.0$.}
    \label{fig:3hdmnfc}
\end{figure}

In Fig.~\ref{fig:3hdmnfc}, we show the constraints in the $\xi_U-\xi_D$ plane for various values of $\xi_E$. An allowed region would appear as a blue region that does not overlap with any of the green and red regions. As can be seen, there is such an allowed region for small values of $|\xi_E|$. For a large enough value of $\xi_E$, BR$(H^+\to\tau^+\nu_\tau)$ becomes larger than the experimental upper bound of Eq.~(\ref{eq:atlasbrln}), and the model is no longer viable. This is demonstrated in the $|\xi_E|=1$ plot in Fig.~\ref{fig:3hdmnfc}.

A testable prediction of this scenario is the following. In the 3HDM-NFC model, Eq.~(\ref{eq:cbcs}) leads to
\beq\label{eq:cbcsnfc}
\frac{\Gamma(H^+\to c\bar s)}{\Gamma(H^+\to c\bar b)}=\left(\left|\frac{V_{cs}}{V_{cb}}\right|\frac{m_s}{m_b}\right)^2\times
\frac{1+|(\xi_U/\xi_D)(m_c/m_s)|^2}
{1+|(\xi_U/\xi_D)(m_c/m_b)|^2}\gtrsim 0.22.
\eeqa
Thus, BR$(H^+\to c\bar s)$ cannot be smaller than BR$(H^+\to c\bar b)$ by more than a factor of $4-5$.   

In the aligned 2HDM (A2HDM) \cite{Pich:2009sp}, Eqs.~(\ref{eq:yukawanfc}) and Eq.~(\ref{eq:yaymnfc}) with three independent parameters hold, similar to the 3HDM model discussed above. (For some subtle differences between the two models, see Ref.~\cite{Cree:2011uy}.) Therefore, the same conclusions apply to A2HDM as well.

\subsection{MFV}
The basic principle of MHDM with minimal flavor violation (MFV) is that the Yukawa couplings of the SM are the only source of breaking of the flavor $[U(3)]^5$ symmetry. In these models, to third order in the Yukawa matrices,
\beqa
\hat Y^U_A&=&(\xi_U+\xi_U^{uu}\hat Y^U_M \hat Y^{U\dagger}_M 
+\xi_U^{dd}V\hat Y^D_M \hat Y^{D\dagger}_M V^\dagger)\hat Y^U_M,\no\\
\hat Y^D_A&=&(\xi_D+\xi_D^{dd}\hat Y^D_M \hat Y^{D\dagger}_M 
+\xi_D^{uu}V^\dagger\hat Y^U_M \hat Y^{U\dagger}_M V)\hat Y^D_M,\no\\
\hat Y^E_A&=&(\xi_E+\xi_E^{ee}\hat Y^E_M \hat Y^{E\dagger}_M)\hat Y^E_M,
\eeqa
where the three $\xi_F$'s and the five $\xi_F^{ff}$'s are dimensionless parameters. We will assume that $|\xi_F^{ff}|\lesssim |\xi_F|$.
The linear terms in $\hat Y^F_M$ have the same structure as in the NFC case, with independent parameters $\xi_U$, $\xi_D$, and $\xi_E$. Thus, their effects are similar to the 3HDM-NFC model of section~\ref{sec:3hdmnfc}. 

The terms $\xi_U^{dd}$ and $\xi_D^{uu}$ introduce a qualitatively new ingredient compared to NFC, which also lead to off-diagonal $\hat Y^F_A$ couplings. The off-diagonal couplings that might be relevant to the $t\to H^+b$ decay are the following:
\beqa
(\hat Y^U_A)_{it}^*V_{ib}/(\hat Y^U_A)_{tt}^*V_{tb}&\approx&(\xi_U^{dd}/\xi_U)y_b^2 |V_{ib}|^2\ \ \ (i=u,c),\no\\
(\hat Y^D_A)_{ib}V_{ti}/(\hat Y^D_A)_{bb}V_{tb}&\approx&(\xi_D^{uu}/\xi_D)y_t^2 |V_{ti}|^2\ \ \ (i=d,s).
\eeqa
We learn that the off-diagonal couplings are small and do not play a significant role in $t\to H^+b$.

The leading off-diagonal couplings that might be relevant to the $H^+\to c\bar b$ decay are the following:
\beqa
(\hat Y^U_A)_{tc}^*V_{tb}/(\hat Y^U_A)_{cc}^*V_{cb}&\approx&(\xi_U^{dd}/\xi_U)y_b^2|V_{tb}|^2,\no\\
(\hat Y^D_A)_{sb}V_{cs}/(\hat Y^D_A)_{bb}V_{cb}&\approx&(\xi_D^{uu}/\xi_D)y_t^2 V_{tb}V_{ts}^*V_{cs}/V_{cb}.
\eeqa
We learn that the $(\hat Y^D_A)_{sb}$ coupling might contribute comparably to $(\hat Y^D_A)_{bb}$ to $H^+\to c\bar b$. In section \ref{sec:hcb} we mentioned that, if $(\hat Y^D_A)_{sb}$ provides the dominant contribution to $H^+\to c\bar b$, we can have ${\rm BR}(H^+\to c\bar b)\simeq1$, provided that $|(\hat Y^D_A)_{ss}|\lesssim 1.3|(\hat Y^D_A)_{sb}|$. In the MFV framework,
\beq
\frac{(\hat Y^D_A)_{ss}}{(\hat Y^D_A)_{sb}}\approx\frac{\xi_D}{\xi_D^{uu}}\frac{y_s}{y_b}\frac{1}{y_t^2 V_{tb}V_{ts}^*}\lesssim 0.4,
\eeq
so the condition is fulfilled.

The leading off-diagonal couplings that might be relevant to the $H^+\to c\bar s$ decays are the following:
\beqa
(\hat Y^U_A)_{tc}^*V_{ts}/(\hat Y^U_A)_{cc}^*V_{cs}&\approx&(\xi_U^{dd}/\xi_U)y_b^2V_{tb}^*V_{ts}V_{cb}/V_{cs},\no\\
(\hat Y^D_A)_{bs}V_{cb}/(\hat Y^D_A)_{ss}V_{cs}&\approx&(\xi_D^{uu}/\xi_D)y_t^2 V_{tb}^*V_{ts}V_{cb}/V_{cs}.
\eeqa
We learn that the off-diagonal couplings are small and do not play a significant role in $H^+\to c\bar s$.

We conclude that in generation-diagonal processes, such as $t\to H^+b$ and $H^+\to c\bar s$, the MFV predictions are similar to the 3HDM-NFC framework, with only small corrections. For generation-off-diagonal processes, such as $H^+\to c\bar b$, there might be ${\cal O}(1)$ deviations from the 3HDM-NFC framework. In any case, the hypothetical signal of Eqs.~(\ref{eq:atlasmh}) and (\ref{eq:atlasbr}) can be accommodated in the MFV framework, in a region of the $(\xi_U,\xi_D,\xi_E)$ parameter space similar to the 3HDM-NFC framework.

\subsection{FN}
\label{sec:fn}
The Froggatt-Nielsen (FN) mechanism explains the smallness and hierarchy of the SM Yukawa couplings~\cite{Froggatt:1978nt} and, furthermore, suppresses the contributions of new physics to flavor-changing processes \cite{Leurer:1992wg}. The basic principle of the FN framework is that of an approximate $U(1)_{\rm FN}$ symmetry under which the fermion generations carry, in general, different charges. The symmetry is explicitly broken by a small spurion field $\epsilon_{\rm FN}$. The FN mechanism is effective in suppressing FCNC in MHDMs if the various Higgs doublets are also charged differently under $U(1)_{\rm FN}$ \cite{Dery:2016fyj}. Phenomenological constraints imply that in the two Higgs doublet models with a FN symmetry (2DHM-FN), $\epsilon_{\rm FN}\lesssim 10^{-3}$ is required. To zeroth order in $\epsilon_{\rm FN}$, the FN mechanism results in NFC-like Yukawa matrices \cite{Dery:2016fyj}. This means that the 2HDM-FN cannot account for Eqs.~(\ref{eq:atlasmh}) and~(\ref{eq:atlasbr}). At first order in $\epsilon_{\rm FN}$, the diagonal entries of $Y_A^F$ deviate from the exact proportionality to $Y_M^F$, and off-diagonal entries appear. For 3HDM with FN symmetry, to first order in $\epsilon_{\rm FN}$, 
\beqa
(\hat Y^F_A)_{ii}&=&\xi_F y_i(1+\epsilon_{\rm FN}\xi_i),\no\\
(\hat Y^F_A)_{ij}&\sim&\epsilon_{\rm FN}\ y_j V_{ij}\ \ (j>i),\no\\
(\hat Y^F_A)_{ij}&\sim&\epsilon_{\rm FN}\ y_j/V_{ji}\ \ (j<i).
\eeqa
The $\sim$ sign here stands for ``up to order one coefficient". The terms of zeroth order in $\epsilon_{\rm FN}$ have the same structure as the NFC case, with three independent parameters. $\xi_U$, $\xi_D$, and $\xi_E$, which are expected to be of ${\cal O}(1)$. Thus, their effects are similar to the 3HDM-NFC model of section~\ref{sec:3hdmnfc}. 

The off-diagonal couplings that might be relevant to the $t\to H^+b$ decay are the following:
\beqa
(\hat Y^U_A)_{ct}^*V_{cb}/(\hat Y^U_A)_{tt}^*V_{tb}&\sim&(\epsilon_{\rm FN}/\xi_U)|V_{cb}|^2/V_{tb},\no\\
(\hat Y^D_A)_{sb}V_{ts}/(\hat Y^D_A)_{bb}V_{tb}&\sim&(\epsilon_{\rm FN}/\xi_D)V_{cb}V_{ts}/V_{tb}.
\eeqa
We learn that the off-diagonal couplings are small and do not play a significant role in $t\to H^+b$.

The leading off-diagonal couplings that might be relevant to the $H^+\to c\bar b$ decay are the following:
\beqa
(\hat Y^U_A)_{tc}^*V_{tb}/(\hat Y^U_A)_{cc}^*V_{cb}&\sim&(\epsilon_{\rm FN}/\xi_U)V_{tb}/|V_{cb}|^2,\no\\
(\hat Y^D_A)_{sb}V_{cs}/(\hat Y^D_A)_{bb}V_{cb}&\sim&(\epsilon_{\rm FN}/\xi_D)V_{cs}.
\eeqa
We learn that, unless $\epsilon_{\rm FN}\ll |V_{cb}|^2$, the $(\hat Y^U_A)_{tc}$ coupling contributes comparably to $(\hat Y^U_A)_{cc}$ to $H^+\to c\bar b$. In section \ref{sec:hcb} we mentioned that $(\hat Y^U_A)_{cc}$ must not give the dominant contribution to $H^+\to c\bar b$ (which is indeed the case as long as $\xi_U\not\gg\xi_D$), so also $(\hat Y^U_A)_{tc}$ does not dominate this decay.

The leading off-diagonal couplings that might be relevant to the $H^+\to c\bar s$ decays are the following:
\beqa
(\hat Y^U_A)_{tc}^*V_{ts}/(\hat Y^U_A)_{cc}^*V_{cs}&\sim&(\epsilon_{\rm FN}/\xi_U)V_{ts}/(V_{cb}V_{cs}),\no\\
(\hat Y^D_A)_{bs}V_{cb}/(\hat Y^D_A)_{ss}V_{cs}&\sim&(\epsilon_{\rm FN}/\xi_D)/V_{cs}.
\eeqa
We learn that the off-diagonal couplings are small and do not play a significant role in $H^+\to c\bar s$.

We conclude that if Eqs.~(\ref{eq:atlasmh}) and (\ref{eq:atlasbr}) are realized in Nature, the 2HDM-FN models will be excluded, while the 3HDM-FN gives predictions similar to the 3HDM-NFC framework.

\section{Minimal scenarios}
\label{sec:minimal}

\subsection{$\boldsymbol{(\hat Y^D_A)_{bb}}$: The minimal scenario}
As pointed out in Section~\ref{sec:hcb}, a minimal scenario that accounts for the hypothetical signal of Eqs.~(\ref{eq:atlasmh}) and (\ref{eq:atlasbr}) is one where a single $\hat Y^F_A$ coupling -- the $(\hat Y^D_A)_{bb}$ coupling -- dominates both $t\to H^+b$ and $H^+\to c\bar b$, while all other $\hat Y^F_A$ couplings are negligibly small (or even vanish). In this case, ${\rm BR}(t\to H^+b)=(1.6\pm0.6)\times10^{-3}$, while ${\rm BR}(H^+\to c\bar b)\simeq1$. The required range is
\beq
0.067\lesssim (\hat Y^D_A)_{bb}\lesssim 0.10,
\eeq
or, equivalently,
\beq\label{eq:minxib}
4\lesssim \xi_b\lesssim 6.
\eeq
This scenario can be described as a ``nightmare scenario", because it demonstrates that there may be no other unavoidable predictions that can be observed in experiments. 

The charged Higgs boson couples to three quark pairs:
\beq
{\cal L}_{H^{\pm}}=-(\overline{t_L}V_{tb}+\overline{c_L}V_{cb}+\overline{u_L}V_{ub})H^+ (\hat Y^D_A)_{bb}b_R+{\rm h.c.}, 
\eeq
with the following consequences:
\begin{itemize}
\item The charmless $H^+$ decay has a very small branching ratio:
\beq
{\rm BR}(H^+\to u\bar b)\simeq|V_{ub}/V_{cb}|^2\sim0.01.
\eeq
\item The $H^+$ contribution to $D^0-\overline{D}{}^0$ mixing is of order $\Delta m_D^{H^+}/\Gamma_D\sim10^{-9}-10^{-8}$, more than five orders of magnitude below the experimental value.
\item With $\xi_{t,c,u}\simeq0$, there is no contribution to the radiative decays $b\to s\gamma$, $t\to c\gamma$ and $c\to u\gamma$.
\end{itemize}

Other implications concern the neutral scalars. In particular, the only tree-level decay mode of the heavy neutral CP-odd scalar $A$ is into $b\bar b$. Given Eq.~(\ref{eq:yayh}), the dominant decay mode of the heavy neutral CP-even scalar $H$ is also into $b\bar b$. The dominant production mode for both $A$ and $H$ is gluon-gluon fusion (ggF), with bottom loop and a cross section proportional to $|(\hat Y^D_A)_{bb}|^2$. 

A search for $A$ and $H$ produced by ggF and decaying into the final $b\bar b$ state has been carried out by the CMS experiment \cite{CMS:2018pwl}. When interpreting their results in the framework of the minimal scenario, one has to take into account the suppression of the production cross section by $|\xi_b y_b/y_t|^2\sim0.005-0.01$ compared to the CMS input parameters, yielding no relevant bounds on $m_{A,H}$.

At tree level, the model predicts $b\bar bA$ and $b\bar bH$ production. A search for heavy neutral scalars produced in association with one or two $b$ quarks and decaying to $b$ quark pairs was carried out by the ATLAS experiment \cite{ATLAS:2019tpq}. In the range $m_{A,H}=450-950$ GeV, upper bounds on $\xi_b$ of order $20-50$ are obtained, well above the required value of Eq.~(\ref{eq:minxib}).

\subsection{$\boldsymbol{(\hat Y^U_A)_{tt}}$ and $\boldsymbol{(\hat Y^U_A)_{tc}}$: The next-to-minimal scenario}
Another scenario involves two $\hat Y^U_A$ couplings: the $(\hat Y^U_A)_{tt}$ coupling accounts for $t\to H^+b$, and the $(\hat Y^U_A)_{tc}$ dominates $H^+\to c\bar b$. In this case, again, ${\rm BR}(t\to H^+b)=(1.6\pm0.6)\times10^{-3}$, while ${\rm BR}(H^+\to c\bar b)\simeq1$. The required range is
\beq
0.067\lesssim (\hat Y^U_A)_{tt}\lesssim 0.10,
\eeq
or, equivalently,
\beq\label{eq:xitntm}
0.067\lesssim \xi_t\lesssim 0.1.
\eeq
The $(\hat Y^U_A)_{tc}$ coupling can assume any value (large enough to make $H^+$ unstable on the collider scale), as long as all other couplings are negligibly small ({\it e.g.}, $(\hat Y^E_A)_{j\tau}\ll0.1$). 

The charged Higgs couples to six quark pairs:
\beq
{\cal L}_{H^{\pm}}=-(\overline{b_L}V_{tb}^*+\overline{s_L}V_{ts}^*+\overline{d_L}V_{td}^*)H^+ [(\hat Y^U_A)_{tt}t_R+(\hat Y^U_A)_{tc}c_R]
+{\rm h.c.},
\eeq
with the following consequences:
\begin{itemize}
\item The bottomless $H^+$ decay has a very small branching ratio:
\beq
{\rm BR}(H^+\to c\bar s)\simeq|V_{ts}/V_{tb}|^2\sim2.5\times10^{-3}.
\eeq
\item The bottomless top decay into $H^+$ has a negligibly small branching ratio:
\beq
{\rm BR}(t\to H^+s)\simeq|V_{ts}/V_{tb}|^2{\rm BR}(t\to H^+b)\sim5\times10^{-6}.
\eeq
\item The $H^+$ contribution to $B^0-\overline{B}{}^0$ mixing is of order $\Delta m_B^{H^+}/\Gamma_B\sim0.013-0.03$, a factor of a few below the current upper bound.
\item With $\xi_{b,s,d}\simeq0$, there is no contribution to the radiative decays $b\to s\gamma$, $t\to c\gamma$ and $c\to u\gamma$.
\end{itemize}

Other implications concern the neutral scalars. In particular, the only tree-level decay modes of the heavy neutral CP-odd scalar $A$ are into $t\bar t$ and $t\bar c+c\bar t$. Given Eq. (\ref{eq:yayh}), the dominant decay modes of the heavy neutral CP-even scalar $H$ are the same. 

The dominant production mode for both $A$ and $H$ is gluon-gluon fusion (ggF), with top loop and a cross section proportional to $|(\hat Y^U_A)_{tt}|^2$. Given Eq.~(\ref{eq:xitntm}), the production cross section for $m_A\sim m_{H^+}$ is about two orders of magnitude below the ggF production cross section of the Higgs boson $h$.

A search for $A$ and $H$ produced by ggF and decaying into final $t\bar t$ state was carried out by the CMS~\cite{CMS:2019pzc} and ATLAS \cite{ATLAS:2017snw} experiments. In the range $m_{A,H}=400-750$ GeV, upper bounds on $\xi_t$ of order $0.6-1$ are obtained, well above the required value of Eq.~(\ref{eq:xitntm}).

At tree level, the model predicts $t\bar tA$ and $t\bar tH$ production. Searches for $t\bar tt\bar t$ events, interpreted in the 2HDM framework, were carried out by the ATLAS~\cite{ATLAS:2022rws} and CMS \cite{CMS:2019rvj} experiments. In the range $m_{A,H}=0.4-1$ TeV, upper bounds on $\xi_t$ of order $0.7-1.6$ are obtained, well above the required value of order $0.1$.

Note that, if $m_t+m_c<m_{A,H}<2m_t$, then the only available tree-level decay mode is $t\bar c+c\bar t$.

If $m_A<m_t$ (and similarly if $m_H<m_t$), then a new decay mode for the top quark, $t\to Ac$, opens up, with 
\beq
\frac{{\rm BR}(t\to Ac)}{{\rm BR}(t\to H^+b)}\approx \left|\frac{(\hat Y^U_A)_{tc}}{(\hat Y^U_A)_{tt}}\right|^2\left(\frac{1-x_{H^+t}}{1-x_{At}}\right)^2.
\eeq
Establishing $t\to H^+b$ and not observing $t\to Ac$ will put constraints in the $(m_A,(\hat Y^U_A)_{tc})$ plane. Note that searches for $t\to qX$, where $X$ decays via $X\to b\bar b$~\cite{ATLAS:2023mcc}, do not apply in this next-to-minimal scenario, because $A$ does not decay into $b\bar b$.

Moreover, from Eq.~(\ref{eq:yayh}) we learn that there is also an off-diagonal Higgs coupling,
\beq
(Y_h)_{tc}=c_{\alpha-\beta}(\hat Y^U_A)_{tc},
\eeq
leading to $t\to ch$ decay. The experimental upper bounds on this mode \cite{CMS:2021hug,ATLAS:2022gzn},
\beq
{\rm BR}(t\to ch)<\left\{
\begin{matrix}9.4\times10^{-4} & {\rm ATLAS},\\
7.3\times10^{-4} & {\rm CMS}.\end{matrix}\right.,
\eeq
put an upper bound, 
\beq
c_{\alpha-\beta}(\hat Y^U_A)_{tc}\lesssim 0.03.
\eeq

\subsection{Bounds from consistency between $m_f$ and $(\hat Y^F_h)_{ff}$}
The SM predicts a strong relation between the fermion masses and their coupling to the Higgs boson, which may break in multi-Higgs doublet models. We define
\begin{equation}
\kappa_f\equiv(\hat{Y}_{h}^{F})_{ff}/y_f.
\end{equation}
The SM predicts $\kappa_f=1$, while 2HDMs predict
\begin{align}
\kappa_f=s_{\beta-\alpha}+c_{\beta-\alpha}\ \xi_f,
\end{align}
where we used Eqs. \eqref{eq:yayh} and \eqref{eq:defxif}. Experiments measuring the SM Higgs production and decay rates constrain $\kappa_f$. 

Since experiments show, in agreement with the SM, that $c_{\beta-\alpha}\ll1$, we approximate $s_{\beta-\alpha}\simeq1$, and obtain:
\beq
c_{\beta-\alpha}\ \xi_f\simeq\kappa_f-1\,.
\eeq
The $1\sigma$ upper bounds on the modifiers of the Yukawa couplings of the SM Higgs boson \cite{ATLAS:2021vrm} provide the following bounds on our parameter space:
\begin{align}
-0.30 & \leq c_{\beta-\alpha}\ \xi_{b}\leq -0.08,\nonumber\\
-0.18 & \leq c_{\beta-\alpha}\ \xi_{t}\leq +0.02\,.
\end{align}
If, in the future, experiments establish $c_{\beta-\alpha}\neq0$, then these bounds will have direct implications for our various scenarios.

\section{Discussion and Conclusions}
\label{sec:con}
The slight excess observed by the ATLAS experiment \cite{ATLAS:2023bzb} in their search for charged Higgs produced in the $t\to H^+b$ decay and decaying via $H^+\to c\bar b$, at $m_{H^+}\sim130$ GeV, motivates us to study the implications of such a light charged Higgs in three directions:
\begin{itemize}
    \item Can such a light $H^+$ be accommodated in multi-Higgs doublet models where the absence of deviations from the SM predictions for flavor-changing processes is naturally explained? In particular, we consider natural flavor conservation (NFC), minimal flavor violation (MFV), and the Froggatt-Nielsen (FN) mechanism as the organizing flavor principles.
    \item Can such a light $H^+$ be accommodated in multi-Higgs doublet models where the flavor structure does not mimic the SM one? In particular, we consider minimal scenarios where one or two new Yukawa couplings account for the signal.
    \item If the signal is established, are there other consequences that unavoidably follow in each viable scenario, which can be used to test the scenario?
\end{itemize}

The first question has previously been studied in Ref.~\cite{Akeroyd:2022ouy}. They focused on NFC models. Where our study overlaps the one in Ref.~\cite{Akeroyd:2022ouy}, we confirm their results. In particular, we confirm that two Higgs doublet models (2HDMs) with NFC cannot explain the signal, while a three-Higgs doublet model (3HDM), where each fermion sector couples to a different Higgs doublet, can do so. Some quantitative differences in this context arise because we use updated experimental constraints on the $b\to s\gamma$ decay, which are considerably stronger than the ones used in Ref.~\cite{Akeroyd:2022ouy}.

In addition to 2HDM-NFC, we find that 2HDM subject to the MFV or to the FN structure will also be excluded if the signal is established. 

The most minimal scenario that can account for the signal has a single coupling, $(\hat Y_A^D)_{bb}$, which dominates both the $t\to H^+b$ and the $H^+\to c\bar b$ decays. The required range for this coupling is given by
\beq
(\hat Y^D_A)_{bb}=(4-6)y_b,
\eeq
where $y_b$ is the SM Yukawa coupling of the bottom quark.

The next-to-minimal scenario that can account for the signal has one coupling, $(\hat Y_A^U)_{tt}$, that dominates the $t\to H^+b$  decay, and another coupling, $(\hat Y^U_A)_{tc}$, which dominates the $H^+\to c\bar b$ decay. The required range for the former is given by
\beq
(\hat Y^U_A)_{tt}=(0.067-0.1)y_t,
\eeq
where $y_t$ is the SM Yukawa coupling of the top quark, while $(\hat Y_A^U)_{tc}$ can assume any value (up to weak constraints from the $H^+$ lifetime and from the non-observation of $t\to hc$).

Further experimental tests of the three viable scenarios that we considered are the following:
\begin{itemize}
    \item For the 3HDM-NFC model, there is a lower bound on the rate of the $H^+\to c\bar s$ decay:
    \beq
    \Gamma(H^+\to c\bar s)/\Gamma(H^+\to c\bar b)\gtrsim 0.22.
    \eeq
    \item For the minimal scenario, the heavy neutral scalars should be produced via $pp\to b\bar bA$ and $pp\to b\bar bH$ at rates which could reach (at most) a factor of order 10 below current bounds. The effects on FCNC processes, such as $D^0-\overline{D}{}^0$ mixing and $t\to c\gamma$ decay, are negligibly small.
    \item For the next-to-minimal scenario, heavy neutral scalars should be produced via $pp\to t\bar tA$ and $pp\to t\bar tH$ at rates that could be (at most) a factor of 50 below current bounds. For $m_{A,H}$ within the range $(m_t+m_c,2m_t)$, $A$ and $H$ appear as $t\bar c$ resonances. For $m_{A,H}<m_t$, the top decays $t\to(A,H)c$ can be observable. Finally, the charged Higgs contribution to $B^0-\overline{B}{}^0$ mixing is non-negligible and can be signaled by a deviation of order $2-5$ percent from the SM prediction for $\Delta m_B$. 
\end{itemize}

\section*{Acknowledgements}
YN is the Amos de-Shalit chair of theoretical physics, and is supported by grants from the Israel Science Foundation (grant number 1124/20), the United States-Israel Binational Science Foundation (BSF), Jerusalem, Israel (grant number 2018257), by the Minerva Foundation (with funding from the Federal Ministry for Education and Research), and by the Yeda-Sela (YeS) Center for Basic Research. 
NB received funding from the Spanish FEDER / MCIU-AEI under the grant FPA2017-84543-P.


\end{document}